\documentclass[runningheads]{llncs}
\usepackage{algorithm,algorithmic}% http://ctan.org/pkg/algorithms
\usepackage{graphicx}
\usepackage{todonotes}
\usepackage{subcaption}
\usepackage{soul}
\usepackage{hyperref}
\usepackage[inline]{enumitem}
\newlist{inlist}{enumerate*}{1}
\setlist[inlist]{label=\textit{(\roman*)}}

\begin{document}
\title{``Knock knock! Who's there?''\\ A study on scholarly repositories' availability}

\titlerunning{``Knock knock! Who's there?'' A study on scholarly repositories' availability}
\author{
    Andrea Mannocci\inst{1}\orcidID{0000-0002-5193-7851} \and
    Miriam Baglioni\inst{1}\orcidID{0000-0002-2273-9004} \and
    Paolo Manghi\inst{1,2}\orcidID{0000-0001-7291-3210}
}

\authorrunning{A. Mannocci et al.}
\institute{CNR-ISTI -- National Research Council, Institute of Information Science and Technologies
``Alessandro Faedo'', 56124 Pisa, Italy\\
\email{name.surname@isti.cnr.it} 
\and
OpenAIRE AMKE, Athens, Greece}
\maketitle              % typeset the header of the contribution
\begin{abstract}
Scholarly repositories are the cornerstone of modern open science, and their availability is vital for enacting its practices.
To this end, scholarly registries such as FAIRsharing, re3data, OpenDOAR and ROAR give them presence and visibility across different research communities, disciplines, and applications by assigning an identifier and persisting their profiles with summary metadata.
Alas, like any other resource available on the Web, scholarly repositories, be they tailored for literature, software or data, are quite dynamic and can be frequently changed, moved, merged or discontinued. Therefore, their references are prone to link rot over time, and their availability often boils down to whether the homepage URLs indicated in authoritative repository profiles within scholarly registries respond or not.

For this study, we harvested the content of four prominent scholarly registries and resolved over 13 thousand unique repository URLs.
By performing a quantitative analysis on such an extensive collection of repositories, this paper aims to provide a global snapshot of their availability, which bewilderingly is far from granted.

\end{abstract}
\section{Introduction}
Scholarly repositories are a vital part of the scholarly infrastructure and therefore are a cornerstone of modern open science practices.
Scholarly registries, such as FAIRsharing\footnote{FAIRsharing -- \url{https://fairsharing.org}}~\cite{sansone2019}, re3data\footnote{re3data -- \url{https://re3data.org}}~\cite{pampel2013}, OpenDOAR\footnote{OpenDOAR -- \url{https://v2.sherpa.ac.uk/opendoar}}, and ROAR\footnote{ROAR -- \url{http://roar.eprints.org/information.html}}, facilitate the discovery and referencing of scholarly repositories by assigning them an identifier (either local or persistent, as for FAIRsharing and re3data) and maintaining a public profile displaying summary metadata, as in \url{https://fairsharing.org/FAIRsharing.wy4egf}.

However, like any other resource on the Web~\cite{bar-yossef2004,cho2003}, scholarly repositories as well, be they tailored for literature, software or data, are dynamic and often changed, moved, merged or discontinued.
Indeed, as scholars soon got to understand, referencing consistently scholarly resources is far from straightforward~\cite{klein2020,klein2014,jones2016a,lawrence2001}.
Therefore, references to repositories are vulnerable to link rot over time, and their availability often boils down, especially for research infrastructures and scholarly communication services aggregating content from registries, to whether their homepages are live and able to respond or not.

In this paper, we harvested the content of four prominent scholarly registries, namely FAIRsharing, re3data, OpenDOAR, and ROAR, and distilled a comprehensive, longitudinal, repository-type-agnostic collection of over 13 thousand unique URLs pointing to scholarly repositories worldwide.
Then, each URL has been requested via HTTP with multiple methods and the response codes, redirection lists and ultimately resolved URLs have been tracked, thus providing a global snapshot of the availability of scholarly repositories.

The analysis highlights that about the 25\% of repository URLs and homepages registered in scholarly registries are problematic.
Even more so, the requests terminating successfully from a syntactical standpoint can still be incorrect semantically, as the content served has nothing to relate with the original scope of the repository.
In our opinion, such results are bewildering, as the URLs considered in this analysis are those available in official repository profiles maintained by scholarly registries and not ``vanilla'' repository references parsed ``in the wild'' from the research literature.

\section{Related work}
Web reference and link rot inside and outside the scholarly domain have been extensively studied over the years~\cite{bar-yossef2004,cho2003,klein2020,klein2022,klein2014,jones2016a,lawrence2001}.
However, to the best of our knowledge, no prior study focused on the availability of scholarly repositories by examining the URLs contained in repository profiles registered by repository managers into scholarly registries.

Despite registries are aware of such issues, and they best-effort notify their users (e.g., \url{https://fairsharing.org/1724} or \url{https://www.re3data.org/repository/r3d100011299}), in several cases, repository profiles could be not yet flagged as problematic.
Furthermore, the authoritative nature of such URLs can open to implications regarding repository management (and not just referencing) that are worth investigating.

In this work, we took inspiration from the methodology introduced in~\cite{klein2020} for studying the persistence of DOIs, which we adapted to the peculiarities of the case study at hand.

\section{Data and methods}
\label{sec:datamethods}
For this analysis, we selected four major scholarly repository registries, namely FAIRsharing, re3data, OpenDOAR and ROAR, whose details are briefly summarised in Table~\ref{tab:data}.
The content of the four registries was dumped by various means in February 2022. 
Each repository profile in the four registries was processed, and its homepage accrued in a list of 13,356 unique URLs.
\begin{table}[t]
\centering
\begin{tabular}{l|l|l|l|l}
Registry    & Dump date & Dump method     & Registry Licence     & \# repos  \\ \hline
FAIRsharing & Feb 2022  & JSON (rest API) & CC-BY-SA    & 1,853        \\
re3data     & Feb 2022  & OpenAIRE        & CC-BY       & 2,793       \\
OpenDOAR    & Feb 2022  & OpenAIRE        & CC-BY-NC-ND & 5,811        \\
ROAR        & Feb 2022  & CSV (website)   & CC-BY       & 5,444     
\end{tabular}
\caption{Overview of the four registries considered in this analysis.}
\label{tab:data}
\vspace{-.5cm}
\end{table}

Each URL was requested via HTTP transactions, as documented in RFC 7231~\cite{fielding2014hypertext}, comprehending a client issuing an HTTP request consisting of a request method and request headers, and a server replying to such request with response headers and (optionally) a response body.
In this study, the HTTP requests adhere to the two most common HTTP request methods, i.e., GET and HEAD. 
As documented in the RFC, the main difference between HEAD and GET is that the server returns a resource representation in the response body for GET requests, which is instead omitted for HEAD requests.
As per RFC 7231, we expect identical results for the two request methods, except for negligible differences in latency.
HTTP requests were issued via Python, were allowed to follow a maximum of 30 redirects, and had a timeout set to 30 seconds.
For each HTTP request successfully concluded, a number of response headers was tracked, namely
\begin{inlist}
    \item original URL;
    \item final URL (same as the original or redirected);
    \item final status code;
    \item redirection chain (in case of redirects);
    \item redirection status codes (in case of redirects);
    \item latency.
\end{inlist}
If the server instead could not respond within this interval, the request was marked as unsuccessful, and the error message (e.g., timeout, connection refused) was noted.
It is worth mentioning that we did not alter the protocol in any way, be it \texttt{http://} or \texttt{https://}, nor did we alter any other part of the original URLs such as ports, as doing so could alter the perceived error rate, despite reducing further the URLs space to process.

For the sake of open science and reproducibility, the content of the registries, the URL list, the code issuing the requests, the collected data, and the Jupyter notebooks for the analysis of the results are available on Zenodo~\cite{andrea_mannocci_2022_6906885}.

\section{Results}
We start our analysis by examining the last HTTP status code returned by the last accessible link in the redirection chain. 
If no redirection takes place, only one status code is returned.
We aggregated the returned status codes in Table~\ref{tab:final}, where rows indicate HTTP request methods, while columns indicate the number of status code occurrences aggregated per response classes (i.e., 2xx for successful requests, 4xx for client errors, and 5xx for server errors).
\begin{table}[t]
\centering
\begin{tabular}{l|c|c|c|c}
Method      & 2xx       & 4xx       & 5xx       & Err\\ \hline
HEAD        & 74.83\% (9,995)     & 6.11\% (816)       &  0.86\% (115)      & 18.2\% (2,431)\\
GET         & 76.05\% (10,158)    & 5.19\% (694)       & 0.76\% (102)       & 18\% (2,403)\\
\end{tabular}
\caption{Final HTTP status codes.}
\label{tab:final}
\end{table}

The largest part of the performed HTTP requests returned 2xx-class status codes (74.83\% and 76.05\% for HEAD and GET, respectively), while the remaining requests exhibited issues of various kinds.
More specifically, about 18\% of HEAD and GET requests failed due to time out, excessive retries or redirects, connection reset by peer, malformed URLs, and so on.
Furthermore, a considerable amount of 4xx-class status codes was returned (i.e., 6.11\% for HEAD, 5.19\% for GET), while a much smaller yet non-negligible quantity of 5xx-class status codes was collected (i.e., 0.86\% for HEAD, 0.76\% for GET).
Please notice that the 3xx-class is relative to redirection, and therefore no final HTTP status code is expected to belong to this class as all redirections were followed, eventually resolving into a non-3xx status code or an error (e.g., for maxing out the number of redirects allowed, or for a timeout).
A significant divergence across the two request methods cannot be observed, even though the slight difference suggests that requests were served differently based on the method used, which infringes the RFC recommendations.
Indeed, \url{https://idr.openmicroscopy.org} triggers an error due to an exceeding number of redirects for HEAD while returning a 200 OK for GET requests.

In total, it is safe to say that about one-quarter of the performed HTTP requests failed to hit their target and resolve successfully, a result that we found bewildering.
This is particularly true, especially if we consider that HTTP requests resolving into 2xx-class status codes, despite being syntactically correct from the protocol perspective, can still be semantically wrong in relation to the content of the web page served.
Indeed, on manual inspection of randomly sampled URLs, we verified that, despite the syntactic correctness, the page resolved had nothing to do with academia or with the repository/project anymore. 
For example, \url{http://www.dlese.org/library/} (\url{https://v2.sherpa.ac.uk/id/repository/425}) resolves, without further redirects, to a web page suggesting that the domain has not been renewed for its original purposes and that someone else put it back onto the market.
Similarly, \url{http://ejournal.windeng.net} (\url{http://roar.eprints.org/1530/}), together with five more URLs, redirects to \url{http://survey-smiles.com} via HEAD, which ostensibly has nothing to do with the initial purpose of the repository.

This event led us to investigate redirections further, which \textit{per se} do not signal a sure malfunction (e.g., redirect from \texttt{http://} to \texttt{https://}) while can still contribute to semantic incorrectness as they could serve an unrelated page (e.g., the repository has been moved somewhere else and the original URL redirects to an error page).
Redirection happens in about 32\% of the requests, i.e., 4,330 and 4,359 for HEAD and GET requests, respectively.
We also noticed that 378 redirection chains for HEAD requests ended up with a status code other than 2xx-class status codes (338 in the 4xx-class, 40 in the 5xx-class), while, for GET requests, 344 redirections chains shared a similar epilogue (307 in the 4xx-class, 37 in the 5xx-class).

No redirection chain longer than five was observed, except for those maxing out the number of redirections allowed.
This was observed sporadically, eight times for HEAD and once for GET, leaving little room for further speculations.
Most redirections resolved in one hop, and the number of N-long redirection chains quickly decreases as N increases.
We experimentally verified that the most frequent reason for one-hop-long redirection chains was to switch from \texttt{http://} to \texttt{https://}, prepend \texttt{www}, append a trailing slash, or combinations of such and similar alterations.

Incidentally, we noticed that the URL space dramatically shrinks, i.e., from the over 13 thousand unique URLs we started with, excluding the errors, 10,926 of them resolved to 9,331 unique URLs via HEAD (-14.59\%), while 10,954 of them resolved to 9,353 unique URLs via GET (-14.61\%).
More specifically, for HEAD requests, a pool of 3,051 unique repository URLs redirected at least twice onto the same URL, for a total of 1,456 unique URLs after redirection.
For GET requests, we registered similar numbers: 3,060 unique repository URLs redirected at least twice onto one of 1,459 unique URLs after redirection.
As it can be noticed, there is no substantial difference between the two methods, yet the behaviour is not identical as it should be expected as per RFC recommendations. 
This is also confirmed by 146 unique URLs for which HEAD and GET returned two different locations.

By aggregating the final URLs after redirections and counting the number of conflated original URLs, we discovered something interesting.
In several cases, for which unfortunately do not exist systematic identification criteria, redirections hid issues worth commenting on.
For example, 16 distinct URLs\footnote{\url{http://brage.bibsys.no/hia/} \{ks, hig, hiak, politihs, hsf, hive, misjon, hinesna, hvo, hibo, histm, dhh, hint, hibu, bdh\}} redirected to \url{https://www.unit.no/ugyldig-lenke-til-dokument-i-vitenarkiv}.
The latter alerts about the incorrectness of the (original) URLs (i.e., ``You are trying to reach a document via a link that is not valid'') and leaves the user with no further indication of the repository's actual location.

Similarly, 6 distinct URLs\footnote{\url{https://toxnet.nlm.nih.gov/cgi-bin/sis/htmlgen?CCRIS}, \url{https://toxnet.nlm.nih.gov/cgi-bin/sis/htmlgen?HSDB}, \url{https://www.nlm.nih.gov/toxnet/index.html}, \url{https://toxmap.nlm.nih.gov/toxmap/}, \url{https://toxnet.nlm.nih.gov/newtoxnet/tri.htm}, \url{https://toxnet.nlm.nih.gov/newtoxnet/cpdb.htm}} conflated onto \url{https://www.nlm.nih.gov/toxnet/index.html}, where a note claims that ``most of NLM's toxicology information services have been integrated into other NLM products and services'', thus leaving no trace of the previous databases and relative information about the content.
Similarly, \url{http://csdb.glycoscience.ru/help/migrate.html} indicates that two original databases redirecting here have been merged into a new one.

On a slightly different note, 7 distinct and seemingly unrelated, URLs\footnote{\url{http://content.wsulibs.wsu.edu/cdm/}, \url{http://idahohistory.cdmhost.com/cdm/}, \url{http://ccdl.libraries.claremont.edu/collection.php?alias=irw}, \url{http://gettysburg.cdmhost.com}, \url{http://cdm16378.contentdm.oclc.org}, \url{http://content.wsulibs.wsu.edu/}, \url{http://trinity.cdmhost.com/index.php}} all redirected to \url{https://www.oclc.org/url/notfound}, whose domain is held by a company positioned in the digital libraries \& services market segment.

Alas, such problems can manifest for any number of conflated URLs, and making a comprehensive list of such cases would equate to examining all the URLs by hand, which quickly grows unfeasible.

\section{Discussion}
The current study suffers from a few limitations.
Firstly, a one-shot resolution provides a snapshot that institution-wide or nationwide infrastructure outages could temporarily distort.
As of June 2022, we tested this hypothesis by checking Ukrainian repositories (i.e., \texttt{.ua} domain) and verified that 36 out of 146 (24.6\%) were unreachable, of which several are in Kharkiv, east Ukraine.
Similarly, other repositories can have downtime for technical reasons.
% \footnote{\url{https://www.cancerdata.org/content/outage-issues-resolved}}.
Therefore, repeating the experiment over an extended timeframe would help recover from spurious transients, provide a better overview of the actual situation, and put scholarly repositories' availability in a long-term perspective.

Furthermore, we resolved the URLs in this analysis via HEAD and GET only. However, we empirically noticed that in some cases, URLs behave differently when accessed via browser, a behaviour already observed in~\cite{klein2019a}.
Therefore, we could extend the study by including HTTP parameters (e.g., user agent, accepted cookies) and by simulating human-like browser interactions via Selenium\footnote{Selenium WebDriver -- \url{https://www.selenium.dev}}.

Nonetheless, the results show that about one out of four URLs from repository profiles in scholarly registries is problematic.
Moreover, this result is a lower bound, as problems could be related to the content served after a successfully served request. 
In fact, as we came to realise, many HTTP requests returned, both with and without redirection, a web page unrelated to the repository pertaining to the original URL.
The methodology described in~\cite{jones2016a} could help to assess the occurrence of such events; however, this would apply only to the cases for which pre- and post- repository registration snapshots exist.
Unfortunately, an exhaustive estimation of content drift would entail a fully manual inspection of repository URLs and the relevant entries on the registries.

As a final remark, repository and IT infrastructure managers have a role in such dysfunctions and should be aware that changes in the infrastructure or, by any means, in the repository lifecycle should be notified and reflected onto registries to strive for accountability and that an incorrect metadata description of a repository can have an impact on reproducibility and open science practices.

\section*{Acknowledgements}
This work was partially funded by the EC H2020 OpenAIRE-Nexus (Grant agreement 101017452).

\bibliographystyle{splncs04}
\bibliography{biblio}
\end{document}